\documentclass[aps,prc,preprint,amsmath,amssymb,showpacs,superscriptaddress,nofootinbib]{revtex4-1}
\usepackage{CJK}
\usepackage{graphicx}
\usepackage{dcolumn}
\usepackage{bm}
\usepackage{color}
\usepackage{float}
\usepackage{hyperref}

\allowdisplaybreaks[4]

\begin{document}

\title{Tetrahedral shape of $^{110}$Zr from covariant density functional theory in 3D lattice space}
\author{F. F. Xu}
\affiliation{State Key Laboratory of Nuclear Physics and Technology, School of Physics, Peking University, Beijing 100871, China}
\author{B. Li}
\affiliation{State Key Laboratory of Nuclear Physics and Technology, School of Physics, Peking University, Beijing 100871, China}
\author{Z. X. Ren}
\affiliation{Institut f\"ur Kernphysik, Institute for Advanced Simulation and J\"ulich Center for Hadron Physics, Forschungszentrum J\"ulich, D-52425 J\"ulich, Germany}
\affiliation{Helmholtz-Institut f\"ur Strahlen- und Kernphysik and Bethe Center for Theoretical Physics, Universit\"at Bonn, D-53115 Bonn, Germany}
\affiliation{State Key Laboratory of Nuclear Physics and Technology, School of Physics, Peking University, Beijing 100871, China}
\author{P. W. Zhao}
\email{pwzhao@pku.edu.cn}
\affiliation{State Key Laboratory of Nuclear Physics and Technology, School of Physics, Peking University, Beijing 100871, China}

\date{\today}
\begin{abstract}
Covariant density functional theory is solved in 3D lattice space by implementing the preconditioned conjugate gradient method with a filtering function (PCG-F).
It considerably improves the computational efficiency compared to the previous inverse Hamiltonian method (IHM).
This new method is then applied to explore the tetrahedral shape of $^{110}$Zr in the full deformation space.
The ground state of $^{110}$Zr is found to have a tetrahedral shape, but the deformations $\beta_{31}$ and $\beta_{33}$ greatly soften the potential energy surface.
This effect is analysed with the microscopic evolution of the single-particle levels near the Fermi surface driven by the deformation.
\end{abstract}
\maketitle

\date{today}

\section{Introduction}\label{sec1}
The occurrence of spontaneous symmetry breaking leads to shapes with a variety of symmetries for nuclei.
Nuclear shape can be described by the parametrization of the nuclear surface $R(\theta,\phi)$ with a multipole expansion \cite{Bohr1975Book},
\begin{eqnarray}
  \label{Eq.q1}R(\theta,\phi)=R_0\left[1+\beta_{00}+\sum_{\lambda=1}^{\infty}\sum_{\mu=-\lambda}^{\lambda}\beta_{\lambda\mu}^{\ast}Y_{\lambda\mu}(\theta,\phi)\right],
\end{eqnarray}
where the $\beta_{\lambda\mu}$'s are the deformation parameters.
The quadrupole shape with axial symmetry characterized by $\beta_{20}$ has been known for a long time, which results in rotational excitation in nuclei \cite{Bohr1975Book}.
In recent decades, many efforts have been devoted to studying the triaxiality ~\cite{Frauendorf1997NPA,Odegard2001PRL,Starosta2001PRL,Meng2006PRC} and reflection asymmetry \cite{Butler1996RMP,Butler2016JPG,Zhao2018IJMPE,Butler2020PRSA} in nuclei, characterized by $\beta_{22}$ and $\beta_{30}$ respectively.
Novel excitation modes have been predicted theoretically to identify these shapes in nuclei \cite{Bohr1975Book,Frauendorf1997NPA,Frauendorf2001RMP,Wang2020SB}, and many of them have been confirmed experimentally \cite{Odegard2001PRL,Starosta2001PRL}.
Indeed, exotic shapes that violate both reflection and axial symmetries, such as tetrahedral shapes, may also exist in nuclei.

A tetrahedral shape corresponds to a finite value of $\beta_{32}$, but vanishing values of all other $\beta_{\lambda\mu}$'s.
The tetrahedral symmetry of nuclei is a direct consequence of the point group $T_{d}^{D}$, which has two two-dimensional and one four-dimensional irreducible representations \cite{Inui1990book}.
Due to the tetrahedral symmetry, the single-particle levels split into multiplets with degeneracies equal to the irreducible representations of the $T_{d}^{D}$ group.
A fourfold degeneracy results in large energy gaps in the single-particle spectrum, and these gaps are comparable to or even larger than the well-known spherical shell gaps.
Empirically, these large gaps occur predominantly in nuclei with $Z(N)=16,20,32,40,56,70$ and $90$, and $N=112,136$, and $142$ \cite{Li1994PRC,Dudek2002PRL,Dudek2007IJMPE,Heiss1999PRC,Arita2014PRC,Dudek2003APPB}.
Thus, a nucleus with proton and/or neutron numbers equal to these values may have a static tetrahedral deformation, characterized by the occurrence of negative-parity bands with missing in-band $E2$ transitions \cite{Bark2010PRL,Jentschel2010PRL}.

Several experiments have been devoted to identifying the tetrahedral shape of nuclei.
The negative-parity bands in $^{160}$Yb and $^{154,156}$Gd have been suggested as candidates for the rotational bands of tetrahedral nuclei \cite{Dudek2006PRL}, but the measured nonzero quadrupole moments contradict the existence of tetrahedral shapes in these nuclei \cite{Bark2010PRL,Doan2010PRC,Jentschel2010PRL}.
For other candidates in nuclei $^{230,232}$U \cite{Ackemann1993NPA}, the possibilities of tetrahedral shapes for the negative-parity bands in $^{230,232}$U appear difficult to reconcile with the systematics of measured quadrupole moments for the neighboring isotone $^{226}$Ra \cite{Ntshangase2010PRC}.
The isomeric state of $^{108}$Zr is proposed to be a candidate for a tetrahedral shape isomer \cite{Sumikama2011PRL}, while the measurement of the corresponding band structure is required to confirm the tetrahedral shape.
The $^{156}$Dy has been suggested as a tetrahedral candidate nucleus \cite{Dobrowolski2011IJMPE}, but it is not supported from the experimental $B(E2)/B(E1)$ ratios of transition probabilities for the negative-parity bands \cite{Hartley2017PRC}.
In conclusion, there is still no firm experimental evidence to support the existence of tetrahedral shapes in nuclei.

The possible tetrahedral shapes in the ground or isomeric states of nuclei have been investigated with many theoretical approaches.
For example, the macroscopic-microscopic (MM) model \cite{Dudek2002PRL,Dudek2007IJMPE,Dudek2006PRL,Schunck2004PRC,Dudek2014PS,Jachimowicz2017PRC,Dudek2018PRC}, the algebraic cluster model \cite{Bijker2014PRL}, the reflection asymmetric shell model \cite{Gao2004CPL,Chen2010NPA}, the nonrelativistic density functional theories (DFTs) \cite{Schunck2004PRC,Takami1998PLB,Yamagami2001NPA,Olbratowski2006IJMPE,Zberecki2006PRC,Zberecki2009PRC,Tagami2013PRC,Tagami2015JPG,Miyahara2018PRC} and the covariant density functional theories (CDFTs) \cite{Zhao2017PRC,Wang2022CTP}.
The CDFT \cite{Meng2016Book} is of particular interest, since it brings many advantages to describe the nuclear systems \cite{Ring1996PPNP,Vretenar2005PR,Meng2006PPNP}, such as the natural inclusion of the self-consistent treatment of the time-odd fields \cite{Meng2013FP} and spin-orbit interactions, which can be clearly seen in the nonrelativistic reduction of the CDFT via a similarity renormalization method \cite{Ren2020PRC}.
However, up to now, the $V_4$ symmetry has always been assumed in the application of CDFT to nuclear tetrahedral shapes \cite{Zhao2017PRC,Wang2022CTP}.

The aim of the present work is to explore the tetrahedral shapes of nuclei in the full deformation space by solving the CDFT in three-dimensional (3D) lattice space.
The CDFT in 3D lattice space has been a long-standing challenge due to the variational collapse \cite{Zhang2010IJMPE} and the Fermion doubling \cite{Tanimura2015PTEP} problems.
It becomes available recently \cite{Ren2019SCI} with the help of the inverse Hamiltonian method (IHM) \cite{Hagino2010PRC} and the Fourier spectral method \cite{Ren2017PRC}.
In Ref.~\cite{Li2020PRC}, a more efficient method, the preconditioned conjugate gradient method with a filtering function (PCG-F), is proposed to solve the nuclear Dirac equation with a given potential in 3D lattice space.
In this work, the CDFT will be solved in 3D lattice space by implementing the PCG-F method \cite{Li2020PRC}, and this new method is then applied to explore the tetrahedral shape of $^{110}$Zr in the full deformation space.
Note that the ground state of $^{110}$Zr was previously predicted to be tetrahedral by the MM model \cite{Schunck2004PRC}, the Skyrme DFTs \cite{Schunck2004PRC,Olbratowski2006IJMPE}, and the multidimensionally constrained CDFT (MDC-CDFT) \cite{Zhao2017PRC}.

The paper is organized as follows, the formulae for the CDFT and the PCG-F method will be briefly introduced in Sec.~\ref{Sec2}.
The numerical details are presented in Sec.~\ref{Sec3}. Section \ref{Sec4} is devoted to the results for tetrahedral shapes in $^{110}$Zr. A summary is given in Sec.~\ref{Sec5}.

\section{Theoretical Framework}\label{Sec2}
\subsection{Formalism of the CDFT}
The starting point of the CDFT is a standard Lagrangian density in the point-coupling form, which can be written as \cite{Zhao2010PRC}
\begin{eqnarray}
  \label{Eq.q2}\mathcal{L}=&&\bar{\psi}(i\gamma^{\mu}\partial_{\mu}-m)\psi\nonumber\\
             &&-\frac{1}{2}\alpha_{S}(\bar{\psi}\psi)(\bar{\psi}\psi)-\frac{1}{2}\alpha_{V}(\bar{\psi}\gamma^{\mu}\psi)(\bar{\psi}\gamma_{\mu}\psi)-\frac{1}{2}\alpha_{TV}(\bar{\psi}\vec{\tau}\gamma^{\mu}\psi)\cdot(\bar{\psi}\vec{\tau}\gamma_{\mu}\psi)\nonumber\\
             &&-\frac{1}{3}\beta_{S}(\bar{\psi}\psi)^3-\frac{1}{4}\gamma_{S}(\bar{\psi}\psi)^4-\frac{1}{4}\gamma_{V}[(\bar{\psi}\gamma^{\mu}\psi)(\bar{\psi}\gamma_{\mu}\psi)]^2\nonumber\\
             &&-\frac{1}{2}\delta_{S}\partial^{\nu}(\bar{\psi}\psi)\partial_{\nu}(\bar{\psi}\psi)-\frac{1}{2}\delta_{V}\partial^{\nu}(\bar{\psi}\gamma^{\mu}\psi)\partial_{\nu}(\bar{\psi}\gamma_{\mu}\psi)-\frac{1}{2}\delta_{TV}\partial^{\nu}(\bar{\psi}\vec{\tau}\gamma^{\mu}\psi)\cdot\partial_{\nu}(\bar{\psi}\vec{\tau}\gamma_{\mu}\psi)\nonumber\\
             &&-\frac{1}{4}F^{\mu\nu}F_{\mu\nu}-e\frac{1-\tau_3}{2}\left(\bar{\psi}\gamma^{\mu}\psi\right)A_{\mu},
\end{eqnarray}
where $m$ is the nucleon mass. According to the conventional variational principle, one obtains the Dirac equation for nucleons,
\begin{eqnarray}
  \label{Eq.q3}\hat{h}(\bm r)\psi_{k}(\bm r)=\left[\bm\alpha\cdot\left(-i\bm\nabla-\bm V(\bm r)\right)+\beta\left(m+S(\bm r)\right)+V^0(\bm r)\right]\psi_{k}(\bm r)=\varepsilon_{k}\psi_{k}(\bm r),
\end{eqnarray}
where $\varepsilon_{k}$ is the single-particle energy.
The single-particle Dirac Hamiltonian $\hat{h}(\bm r)$ contains the scalar $S(\bm r)$ and four-vector $V^{\mu}(\bm r)$ potentials,
\begin{subequations}
  \begin{eqnarray}
    \label{Eq.q4a}S(\bm{r})=&&\alpha_S\rho_S+\beta_S\rho_S^2+\gamma_S\rho_S^3+\delta_S\Delta\rho_S,\\
    \label{Eq.q4b}V^\mu(\bm{r})=&&\alpha_Vj^\mu+\gamma_V(j^\mu j_\mu)j^\mu+\delta_V\Delta j^\mu+\tau_3\alpha_{TV}j_{TV}^\mu+\tau_3\delta_{TV}\Delta j_{TV}^\mu+e\frac{1-\tau_3}{2}A^\mu,
  \end{eqnarray}
\end{subequations}
where the electromagnetic field $A^{\mu}$ is determined by Poisson's equation, and the densities and currents are defined as
\begin{subequations}\label{Eq.q5}
  \begin{eqnarray}
    \rho_{S}(\bm r)=&&\sum_{k}v_{k}^2\bar{\psi}_{k}(\bm r)\psi_{k}(\bm r),\\
    j^{\mu}(\bm r)=&&\sum_{k}v_{k}^2\bar{\psi}_{k}(\bm r)\gamma^{\mu}\psi_{k}(\bm r),\\
    \vec{j}_{TV}^{\mu}(\bm r)=&&\sum_{k}v_{k}^2\bar{\psi}_{k}(\bm r)\gamma^{\mu}\tau_3\psi_{k}(\bm r),\\
    j_{c}^{\mu}(\bm r)=&&\sum_{k}v_{k}^2\bar{\psi}_{k}(\bm r)\gamma^{\mu}\frac{1-\tau_3}{2}\psi_{k}(\bm r).
  \end{eqnarray}
\end{subequations}
Here, $\tau_3$ is the isospin Pauli matrix with the eigenvalues $+1$ for neutrons and $-1$ for protons.
The time component of the vector current $j^{\mu}$ is usually denoted as the vector density $\rho_{v}$.

For open shell nuclei, pairing correlations play an important role, and they are taken into account with the BCS method. The pairing energy functional is given by
\begin{eqnarray}
  \label{Eq.q6}E_{{\rm pair}}=-\sum_{\tau=n,p}\frac{G_{\tau}}{4}\int d^3r\kappa_{\tau}^{*}(\bm r)\kappa_{\tau}(\bm r),
\end{eqnarray}
where $G_{\tau}$ is the constant pairing strength and $\kappa(\bm r)$ is the pairing tensor,
\begin{eqnarray}
  \label{Eq.q7}\kappa(\bm r)=2\sum_{k>0}f_{k}u_{k}v_{k}|\psi_{k}(\bm r)|^2,
\end{eqnarray}
with the smooth-cutoff weight factor
\begin{eqnarray}
  \label{Eq.q8}f_{k}=\frac{\Theta(-\varepsilon_k)}{1+{\rm exp}[(\varepsilon_{k}-\lambda_{{\rm F}}-\Delta E_{\tau})/\mu_{\tau}]}.
\end{eqnarray}
Here, the Fermi energy $\lambda_{{\rm F}}$ is determined by the particle number, $2\sum_{k>0}v_{k}^2=N_{\tau}$, with $N_{\tau}$ the particle number of neutrons or protons.
The cutoff parameters $\Delta E_{\tau}=5$ MeV and $\mu_{\tau}=\Delta E_{\tau}/10=0.5$ MeV are chosen as in Ref.~\cite{Ryssens2015CPC}.
$\Theta(-\varepsilon_k)$ equals one for bound levels and zero elsewhere, and it is introduced to exclude the continuum in the pairing window.

\subsection{Implementation of the PCG-F method}
In the PCG-F method, the lowest $\tilde{A}$ eigenstates in the Fermi sea of the Dirac equation~\eqref{Eq.q3} are solved iteratively starting from a set of orthonormalized guess solutions $\psi_{k}^{(0)}~(k=1,2,...,\tilde{A})$.
Here, the value of $\tilde{A}$ is chosen to include all bound states.
The trial wave function $\psi_k$ is then updated iteratively,
\begin{eqnarray}\label{Eq.q9}
   \psi_k^{(i+1)} = \sum_{l=0}^{\tilde{A}}\left[G_{kl}^aX_l^{(i)} + G_{kl}^bW_l^{(i)} + G_{kl}^cP_l^{(i)}\right]\qquad (i=0,1,2,...),
\end{eqnarray}
where $X_l^{(i)}$, $W_l^{(i)}$, and $P_l^{(i)}$ are defined as,
\begin{subequations}\label{Eq.q10}
  \begin{eqnarray}
      X_l^{(i)}=&&F(\hat{h}^{(i)})\psi_l^{(i)},\\
      W_l^{(i)}=&&F^4(\hat{h}^{(i)})T_l^{(i)}\left[\hat{h}^{(i)}-\langle\psi_l^{(i)}|\hat{h}^{(i)}|\psi_l^{(i)}\rangle \right]\psi_l^{(i)},\\
      P_l^{(i)}=&&F(\hat{h}^{(i)}) \left[\psi_l^{(i)}-\sum_{l'=1}^{\tilde{A}}\langle\psi_{l'}^{(i-1)}|\psi_l^{(i)}\rangle\psi_{l'}^{(i-1)}\right].
  \end{eqnarray}
\end{subequations}
The initial $P_l^{(0)}$ is set to zero.
The filtering operator $F(\hat{h})$ and the preconditioner $T_l$ are introduced for the sake of iteration convergence.
The single-particle Dirac Hamiltonian $\hat{h}^{(i)}$ is constructed from the densities and currents determined by the wave functions $\{\psi_k^{(i)}\}$.
The coefficient matrices $G^a$, $G^b$, and $G^c$ in Eq.~\eqref{Eq.q9} are chosen to minimize $\sum_{k=1}^{\tilde{A}}\langle\psi_k^{(i+1)}|\hat{h}^{(i)}|\psi_k^{(i+1)}\rangle$ under the orthonormalization condition $\langle\psi_k^{(i+1)}|\psi_l^{(i+1)}\rangle=\delta_{kl}$.

Similar to Ref.~\cite{Li2020PRC}, the filtering operator $F(\hat{h})$ and the preconditioner $T_l$ read,
\begin{eqnarray}\label{Eq.q11}
  &&F(\hat{h}^{(i)})=\left(\hat{h}^{(i)}+2m\right)^2,\\
  &&T_l^{(i)} =\left[\hat{p}^2+\left(g_l^{(i)}m\right)^2\right]^{-1},
\end{eqnarray}
with
\begin{eqnarray}\label{Eq.q13}
   g_l^{(i)}=0.15\frac{\langle\psi_l^{(i)}|\hat{h}^{(i)}|\psi_l^{(i)}\rangle}{(V^0+S)_{\rm min}}+0.10.
\end{eqnarray}

There are two criteria for the convergence of the iteration.
One is that the energy dispersions $\langle\psi_l^{(i)}|[\hat{h}^{(i)}]^2|\psi_l^{(i)}\rangle-\langle\psi_l^{(i)}|\hat{h}^{(i)}|\psi_l^{(i)}\rangle^2$ for all occupied levels should be smaller than a certain value, e.g., $10^{-8}~{\rm MeV}^2$.
The other one is that the differences between the mean potentials [Eqs.~\eqref{Eq.q4a} and~\eqref{Eq.q4b}] at two adjacent iterations should be smaller than a certain value.
The convergence is achieved only if both criteria are satisfied.

\subsection{Nuclear bulk properties}
From the converged wave functions, the nuclear total energy and the deformation parameters can be calculated.
The total energy consists of the mean-field energy $E_{{\rm MF}}$, the pairing energy $E_{{\rm pair}}$ and the center-of-mass (c.m.) correction energy $E_{{\rm cm}}$
\begin{eqnarray}
  \label{Eq.q14}E_{{\rm tot}}=E_{{\rm MF}}+E_{{\rm pair}}+E_{{\rm cm}},
\end{eqnarray}
where the mean-field energy $E_{{\rm MF}}$ is written as
\begin{eqnarray}
  \label{Eq.q15}E_{{\rm MF}}=&\displaystyle\int&d^3r\bigg\{\sum_{k}v_{k}^2\psi_{k}^{\dagger}(\bm\alpha\cdot\bm p+\beta m)\psi_{k}\nonumber\\
   &+&\frac{\alpha_{S}}{2}\rho_{S}^2+\frac{\alpha_{V}}{2}j^{\mu}j_{\mu}+\frac{\alpha_{TV}}{2}\vec{j}_{TV}^{\mu}\cdot(\vec{j}_{TV})_{\mu}
   +\frac{\beta_{S}}{3}\rho_{S}^3+\frac{\gamma_{S}}{4}\rho_{S}^4+\frac{\gamma_{V}}{4}\left(j^{\mu}j_{\mu}\right)^2\nonumber\\
   &+&\frac{\delta_{S}}{2}\rho_{S}\Delta\rho_{S}+\frac{\delta_{V}}{2}j^{\mu}\Delta j_{\mu}+\frac{\delta_{TV}}{2}\vec{j}_{TV}^{\mu}\cdot\Delta(\vec{j}_{TV})_{\mu}
   +\frac{1}{2}A_{\mu}\Delta A^{\mu}+ej_{c}^{\mu}A_{\mu}\bigg\}.
\end{eqnarray}
The pairing energy $E_{{\rm pair}}$ is calculated following Eq.~\eqref{Eq.q6}, and the c.m. correction energy $E_{{\rm cm}}$ is considered with the microscopic c.m. correction
\begin{eqnarray}
  \label{Eq.q24}E_{{\rm cm}}=-\frac{1}{2mA}\langle\bm P_{{\rm cm}}^2\rangle,
\end{eqnarray}
with $A$ the mass number and $\bm P_{{\rm cm}}=\sum_{k}\bm p_{k}$ the total momentum in the c.m. frame.

The deformation parameters $\beta_{\lambda\mu}$'s are calculated with
\begin{eqnarray}
  \label{Eq.q25}\beta_{\lambda\mu}=\frac{4\pi}{3AR^{\lambda}}\int d^3r\rho_{v}(\bm r)r^{\lambda}Y_{\lambda\mu},
\end{eqnarray}
where $Y_{\lambda\mu}$ is the spherical harmonics and $R=1.2\times A^{1/3}$ fm.
Note that we need to additionally constrain the center of mass of the whole nucleus at the origin of the 3D box, and align the principal axes with the coordinate axes to remove the redundant degrees of freedom.

\section{Numerical details}\label{Sec3}
In this work, the point-coupling density functional PC-PK1 \cite{Zhao2010PRC} is used.
For the 3D lattice space, the step sizes and the grid numbers along the $x$, $y$ and $z$ axes are chosen as $1$ fm and 30, respectively.
Similar to the Ref. \cite{Zhao2017PRC}, the neutron and proton pairing strengths $G_n=-330~{\rm MeV}\cdot{\rm fm}^3$ and $G_p=-430~{\rm MeV}\cdot{\rm fm}^3$, are determined by reproducing the empirical pairing gaps of $^{102,104}$Zr, which are obtained with the three-point odd-even mass differences formula \cite{Bender2000EPJA} (see Table~\ref{Tab1}).

\begin{table}[!htbp]
  \centering
  \caption{Pairing gaps (in MeV) calculated by the CDFT in 3D lattice space for $^{102,104}$Zr, in comparison with the empirical values extracted from the three-point odd-even mass differences. The experimental masses are taken from AME2020 ~\cite{Wang2021CPC}.}\label{Tab1}
  \begin{ruledtabular}
  \begin{tabular}{ccccc}
                & \multicolumn{2}{c}{$^{102}$Zr}     & \multicolumn{2}{c}{$^{104}$Zr}    \\
                & $\Delta_n$  & $\Delta_p$         & $\Delta_n$  & $\Delta_p$        \\\hline
    \rule{0pt}{8pt}
    Empirical  & 1.10         & 1.54                & 1.08         & 1.53                \\
    CDFT      & 1.12         & 1.55                & 1.00         & 1.49                 \\
  \end{tabular}
  \end{ruledtabular}
\end{table}

\section{Results and discussion}\label{Sec4}

\begin{figure}[!htbp]
  \centering
  \includegraphics[width=0.45\textwidth]{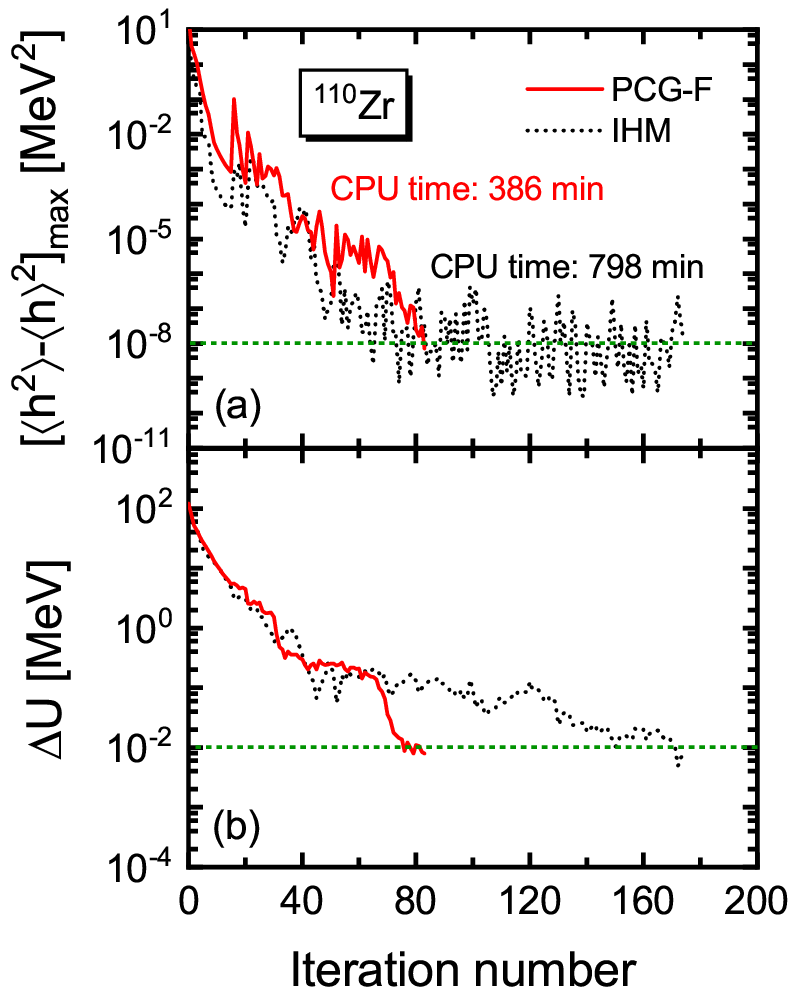}
  \caption{(Color online.) The maximum energy dispersion for the occupied single-particle states (a), and the maximum absolute difference between the mean potentials (including the scalar and vector potentials) at two adjacent iterations (b), as functions of the iteration number for the ground state of $^{110}$Zr.
  The solid and dotted curves respectively represent the results calculated by the PCG-F and IHM, and the corresponding computation times are also given.}
  \label{Fig1}
\end{figure}
We firstly discuss the efficiency of the PCG-F method in the self-consistent CDFT calculations for the ground state of $^{110}$Zr.
In Fig.~\ref{Fig1}, the maximum energy dispersion $\left[\langle h^2\rangle-\langle h\rangle^2\right]_{{\rm max}}$ for the occupied single-particle states and the maximum absolute difference $\Delta U$ between the mean potentials at two adjacent iterations are shown, in comparison with the results given by the IHM.
For the PCG-F method, it takes only 84 iterations to achieve the convergence, i.e., $\left[\langle h^2\rangle-\langle h\rangle^2\right]_{{\rm max}}\leq10^{-8}\ {\rm MeV}^2$ and $\Delta U\leq10^{-2}\ {\rm MeV}$, while it requires more than 170 iterations for the IHM to reach the same accuracy.
The difference between the total energies obtained in these two methods is smaller than $10^{-5}$ MeV.
The total computational time for the PCG-F method is 386 minutes with the Intel(R) Xeon(R) CPU E5-2680, and it saves 52\% of the computational time as compared with the IHM calculations.
As seen in Ref.~\cite{Li2020PRC}, compared with the IHM, the PCG-F method gives a much faster convergence in solving the Dirac equation with a given potential.
The present results prove that the PCG-F method is more efficient than the IHM in the self-consistent CDFT calculations as well.
This is not trivial because, during the self-consistent solution of the CDFT in 3D lattice space, the Dirac equation is not exactly solved until the self-consistency is achieved.
In the following, we apply the framework of the PCG-F method to study the tetrahedral shape of $^{110}$Zr.

\begin{figure}[!htbp]
  \centering
  \includegraphics[width=0.60\textwidth]{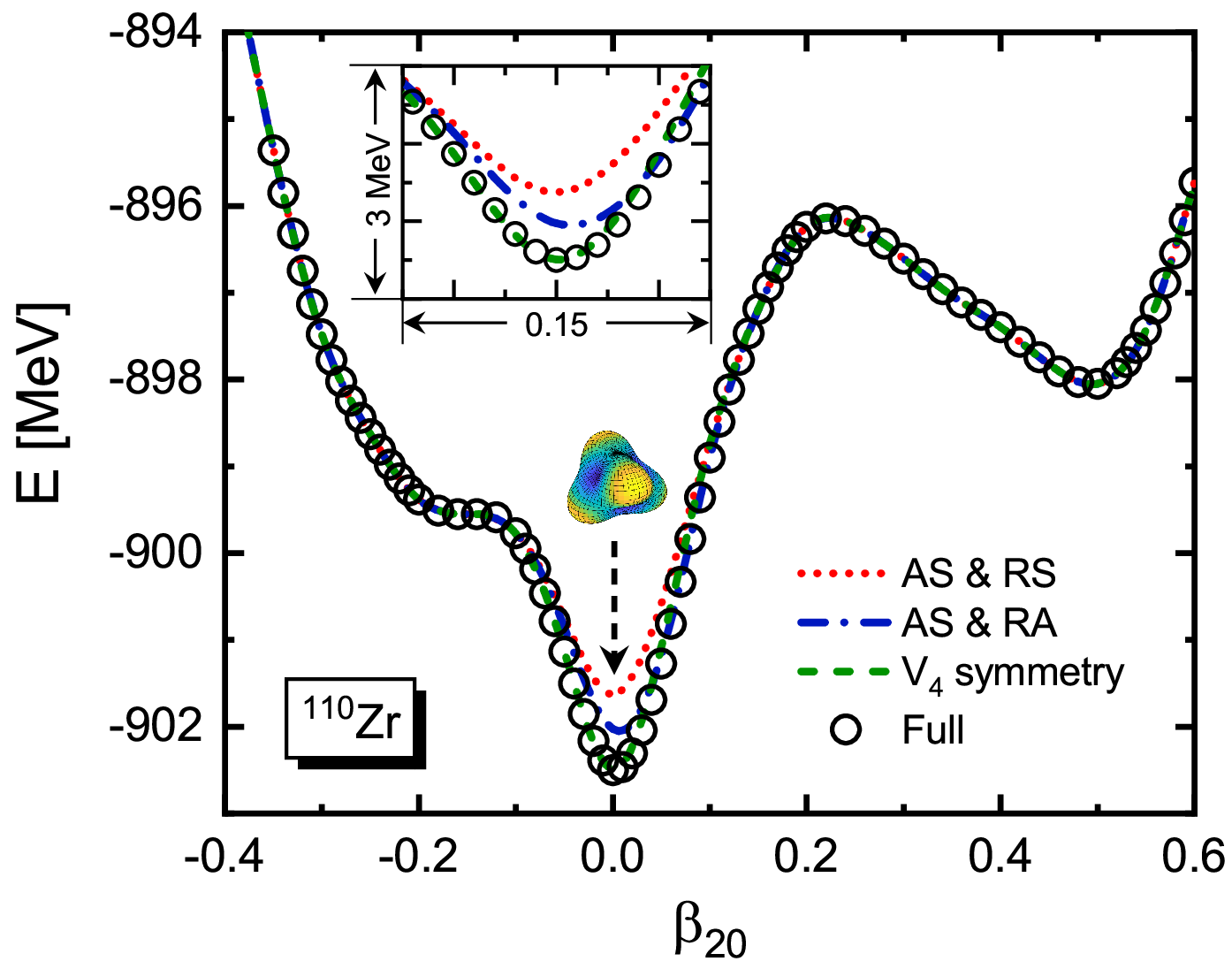}
  \caption{(Color online.) The potential energy curves of $^{110}$Zr calculated with the CDFT in 3D lattice space by imposing different symmetries.
  The results restricted to axial symmetry and reflection symmetry (AS\ \&\ RS), axial symmetry and reflection asymmetry (AS\ \&\ RA), and
  $V_4$ symmetry (including deformations $\beta_{\lambda\mu}$ with even $\mu$) are represented by dotted, dashed-dotted, and dashed lines, respectively.
  The open circles represent the results without any symmetry restriction (Full), and the obtained ground-state shape is illustrated by the 3D image.
  The inset zooms in to the detailed structure of the potential energy curves near $\beta_{20}=0$.}
  \label{Fig2}
\end{figure}
Figure~\ref{Fig2} depicts the one-dimensional potential energy curves of $^{110}$Zr calculated with the CDFT in 3D lattice space by imposing different symmetry restrictions: (i) axial and reflection symmetry (AS\ \&\ RS), (ii) axial symmetry and reflection asymmetry (AS\ \&\ RA), (iii) $V_4$ symmetry (including deformations $\beta_{\lambda\mu}$ with even $\mu$), and (iv) full deformation space including all $\beta_{\lambda\mu}$'s (Full).
There are two energy minima in all cases, i.e., the ground state at $\beta_{20}\approx0.00$ and a prolate minimum at $\beta_{20}\approx0.50$.
The ground-state energy varies visibly in the calculations with different symmetry restrictions.
A spherical ground state is obtained if one assumes axial and reflection symmetry.
The ground-state energy is lowered by about $0.4$ MeV if one releases the restriction of reflection symmetry.
It is lowered further by about $0.5$ MeV if the nonaxial deformation is allowed except the deformation of $\beta_{\lambda\mu}$ with odd $\mu$.
This ground state has a tetrahedral shape ($\beta_{20}\approx0.00,\beta_{30}\approx0.00,\beta_{32}\neq0.00$), and it is consistent with the results obtained in the previous MDC-CDFT calculations \cite{Zhao2017PRC}.

\begin{table}[!htbp]
  \centering
  \caption{The quadrupole deformations $\beta_{20}$ and $\beta_{22}$, octupole deformations $\beta_{30},\beta_{31},\beta_{32}$ and $\beta_{33}$, hexadecapole deformation $\beta_{40}$ and the total energy $E$ for the ground state of $^{110}$Zr calculated by the CDFT in 3D lattice space.}\label{Tab2}
  \begin{ruledtabular}
  \begin{tabular}{cccccccc}
    $\beta_{20}$ & $\beta_{22}$ & $\beta_{30}$ & $\beta_{31}$ & $\beta_{32}$ & $\beta_{33}$ & $\beta_{40}$ & $E$ [MeV]  \\\hline
    \rule{0pt}{8pt}
         0.00    &   0.00       &     0.00     &     0.01     &     0.15     &     0.01     &    -0.02     &  -902.49
  \end{tabular}
  \end{ruledtabular}
\end{table}

Thanks to the solutions in the 3D lattice space, one could remove all symmetry restrictions, and the results are shown by open circles in Fig.~\ref{Fig2}.
The ground-state energy barely changes, and this implies quite small deformations beyond $V_4$ symmetry, like $\beta_{31}$ and $\beta_{33}$.
The obtained ground-state deformation parameters as well as the total energy for $^{110}$Zr are listed in Table~\ref{Tab2}.
The same results are found in the calculations with the same box size but a smaller step size of $0.8$ fm.
From the listed deformations in Table~\ref{Tab2}, one may conclude that the tetrahedral shape still exists in the full deformation space.

\begin{figure}[!htbp]
  \centering
  \includegraphics[width=0.60\textwidth]{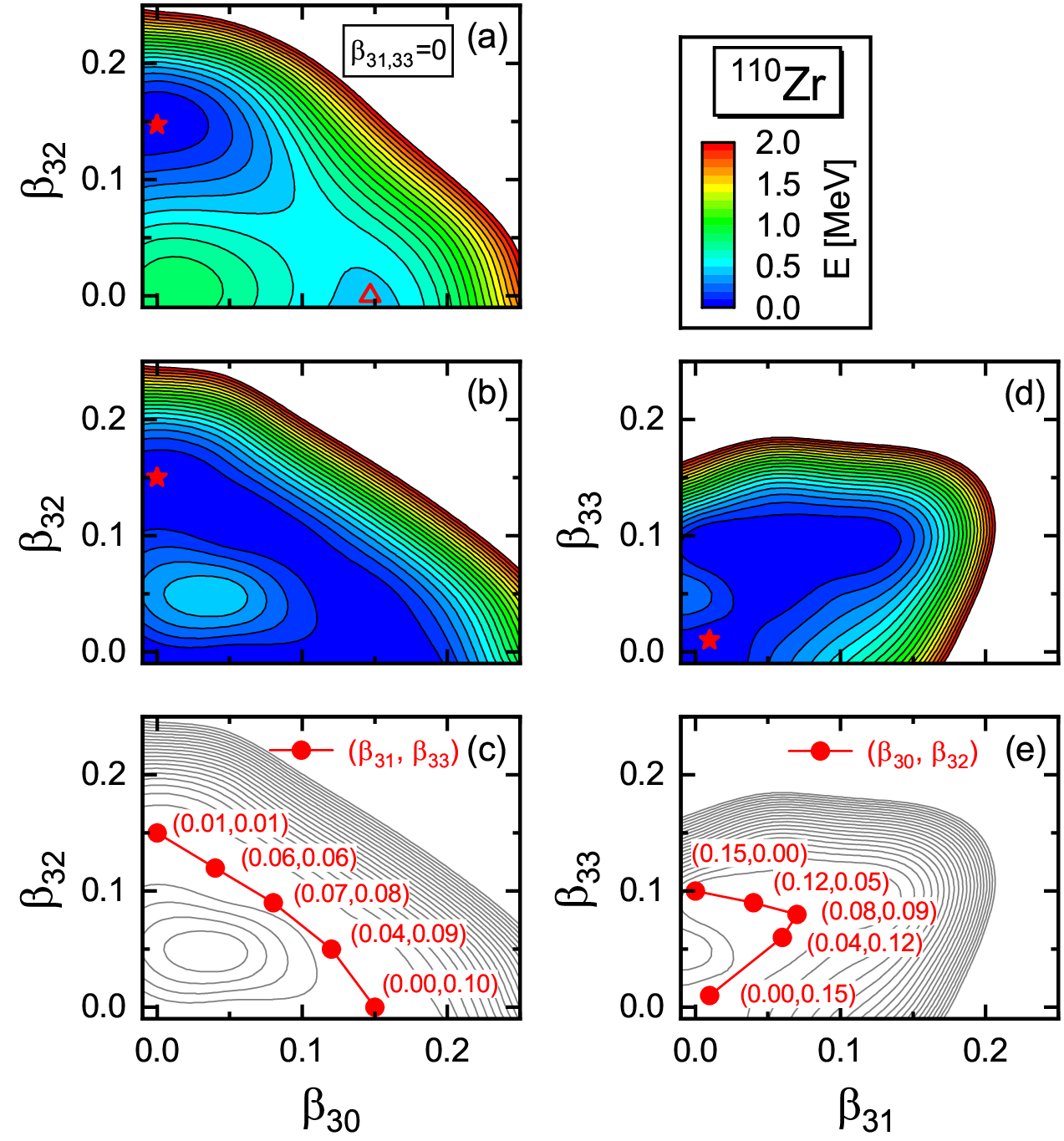}
  \caption{(Color online.) The potential energy surfaces of $^{110}$Zr in the ($\beta_{30}, \beta_{32}$) and ($\beta_{31}, \beta_{33}$) planes.
  The energies are normalized with respect to the ground-state energy, and the contour interval is $0.1$ MeV.
  Note that $\beta_{31}$ and $\beta_{33}$ are always constrained to zero in panel (a), where the tetrahedral minimum and the pear-like minimum are indicated by a solid star and an open triangle, respectively.
  In panel (c), the solid line connects the lowest energy points from $\beta_{30}\approx0.00$ to $\beta_{30}\approx0.15$, and the deformations $(\beta_{31},\beta_{33})$ are marked at some representative points.
  The same points are also shown in panel (e), where the $(\beta_{30},\beta_{32})$ deformations are marked.
  }
  \label{Fig3}
\end{figure}

We further investigate the softness of the tetrahedral shape in the ground state of $^{110}$Zr, against the $\beta_{31}$ and $\beta_{33}$ deformations.
In Fig.~\ref{Fig3}, we show the potential energy surfaces in the ($\beta_{30},\beta_{32}$) and ($\beta_{31},\beta_{33}$) planes.
The quadrupole deformations $\beta_{20}$ and $\beta_{22}$ are determined self-consistently, and their values are found to be very close to zero.
In Fig.~\ref{Fig3}(a), the octupole deformations $\beta_{31}$ and $\beta_{33}$ are constrained to zero.
It shows a well-developed tetrahedral ground state with $(\beta_{30},\beta_{32})\approx(0.00,0.15)$ and a pear-like isomeric state at $(\beta_{30},\beta_{32})\approx(0.15,0.00)$.
The barrier between the two minima is about 0.5 MeV.
After releasing all symmetry restrictions, as seen in Fig.~\ref{Fig3}(b), the tetrahedral ground state remains but the barrier vanishes.
In fact, there is a rather flat path connecting the tetrahedral shape $(\beta_{30},\beta_{32})\approx(0.00,0.15)$ and the pear shape $(\beta_{30},\beta_{32})\approx(0.15,0.00)$ in the potential energy surface.
This path can be seen more clearly in Fig.~\ref{Fig3}(c), and the nonzero $(\beta_{31},\beta_{33})$ values in the parentheses imply the importance of the $\beta_{31}$ and $\beta_{33}$ deformations.
As seen in Fig.~\ref{Fig3}(d), the tetrahedral ground state is also very soft in the $(\beta_{31},\beta_{33})$ plane.
The corresponding lowest-energy path connecting the tetrahedral shape and the pear shape, as seen in Fig.~\ref{Fig3}(e), is also very flat.
All these results prove that the tetrahedral ground state in $^{110}$Zr is greatly softened in the full deformation space.

\begin{figure}[!htbp]
  \centering
  \includegraphics[width=0.5\textwidth]{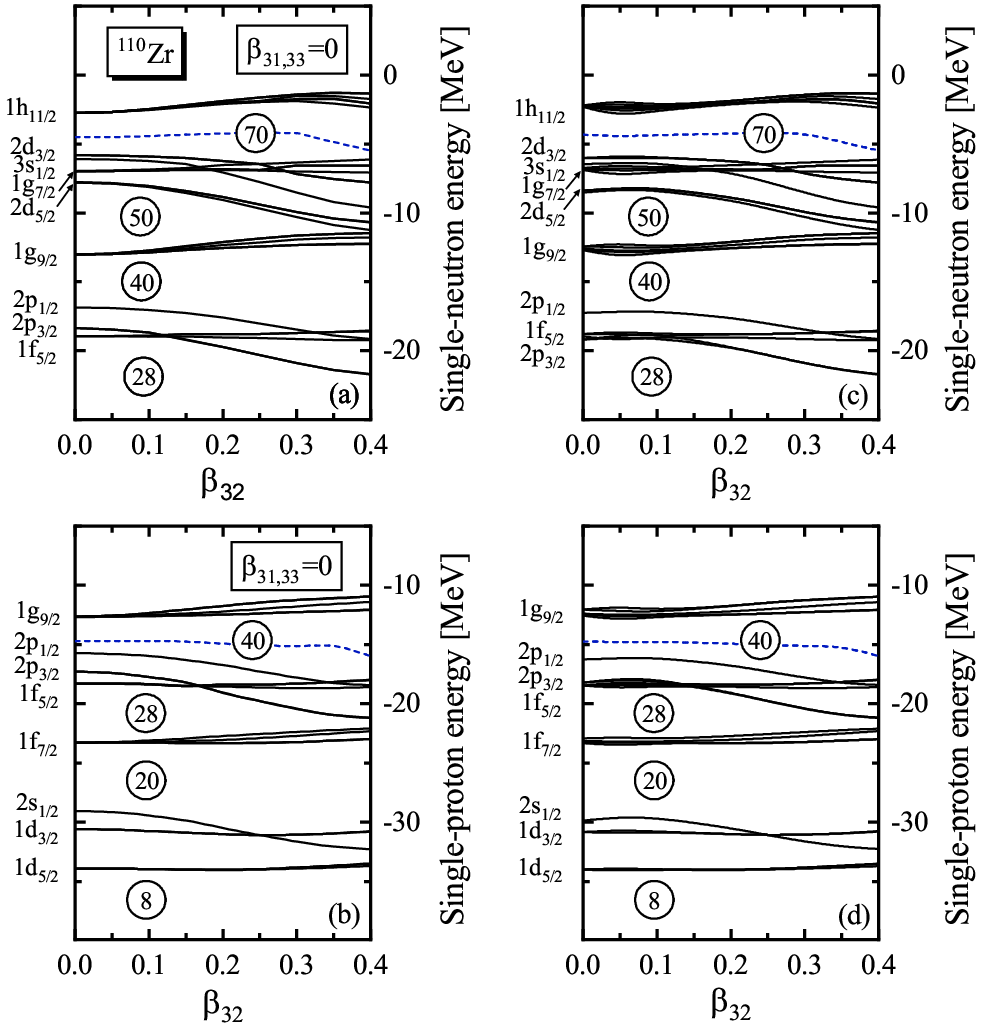}
  \caption{(Color online.) Single-neutron and -proton levels of $^{110}$Zr as functions of $\beta_{32}$. In panels (a) and (b), the $\beta_{31}$ and $\beta_{33}$ are fixed to zero, while for panels (c) and (d), all deformations are allowed.
  The levels are labeled by the corresponding spherical quantum number of their main component.
  The dashed lines represent the Fermi levels.}
  \label{Fig4}
\end{figure}

\begin{figure}[!htbp]
  \centering
  \includegraphics[width=0.5\textwidth]{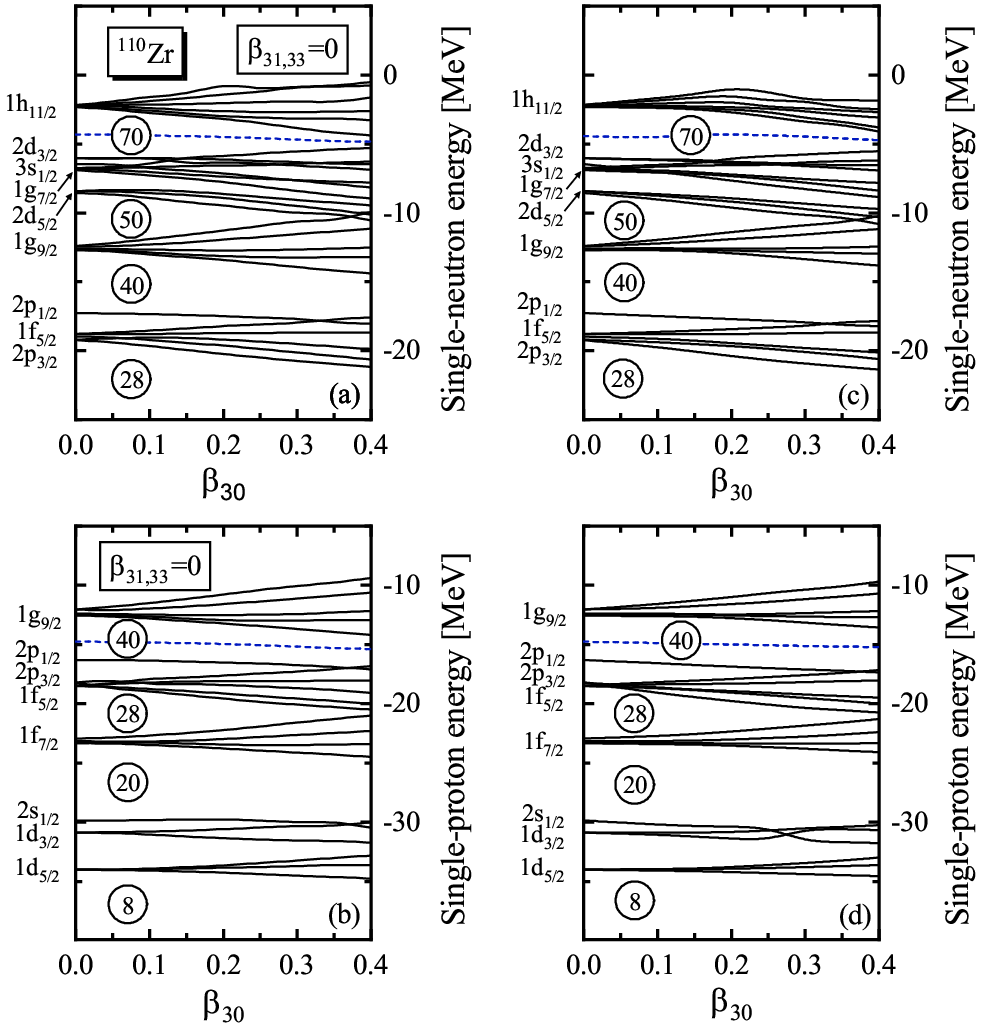}
  \caption{(Color online.) Same as Fig.~\ref{Fig4}, but for the single-particle levels as functions of $\beta_{30}$ and the $\beta_{32}$ is fixed at 0.15.}
  \label{Fig5}
\end{figure}

For a microscopic understanding of the impacts of the $\beta_{31}$ and $\beta_{33}$ deformations on the tetrahedral ground state, the single-particle levels of $^{110}$Zr near the Fermi surface are shown in Figs.~\ref{Fig4} and \ref{Fig5} as functions of $\beta_{32}$ and $\beta_{30}$, respectively.
In Figs.~\ref{Fig4}(a) and \ref{Fig4}(b), the deformations $\beta_{31}$ and $\beta_{33}$ are constrained to zero, and this leads to a vanishing $\beta_{20}$. Therefore, at $\beta_{32}=0$, the single-particle levels are degenerate according to their $j$ values.
With the increasing $\beta_{32}$, the single-particle levels split into multiplets with degeneracies equal to the irreducible representations of the $T_{d}^{D}$ group due to the tetrahedral symmetry. For example, the spherical levels with $j=5/2$ are sixfold degenerate and they can be reduced to the two-dimensional irreducible representation and four-dimensional irreducible representation of the $T_{d}^{D}$ group, and these levels split into two levels with degeneracies 2 and 4 as $\beta_{32}$ increases.
This is consistent with the previous study with the MDC-CDFT in Ref.~\cite{Zhao2017PRC}.
The energy gaps at $N=70$ and $Z=40$ grow gradually with the increasing $\beta_{32}$, and a remarkable $\beta_{32}$ deformation is expected for $^{110}$Zr.
In Figs.~\ref{Fig4}(c) and \ref{Fig4}(d), due to the nonzero $\beta_{31}$ and $\beta_{33}$ values, the spherical symmetry is broken, and the single-particle levels are sightly split even at $\beta_{32}=0$.
The energy gaps at $N=70$ and $Z=40$ are roughly constant with the increasing $\beta_{32}$, and this is associated with the soft tetrahedral ground state in the $\beta_{32}$ direction.

By fixing $\beta_{32}=0.15$, the single-particle levels of $^{110}$Zr are shown in Fig.~\ref{Fig5} as functions of $\beta_{30}$.
In Figs.~\ref{Fig5}(a) and \ref{Fig5}(b), a sharp decline of the energy gaps at $N=70$ and $Z=40$ is found with the increasing $\beta_{30}$, and this leads to a tetrahedral ground state with vanishing $\beta_{30}$ for $^{110}$Zr.
In contrast, the energy gaps at $N=70$ and $Z=40$, as seen in Figs.~\ref{Fig5}(c) and \ref{Fig5}(d), vary only gently with the increasing $\beta_{30}$, and this is reflected by the soft nature of the tetrahedral ground state in the $\beta_{30}$ direction.

\section{Summary}\label{Sec5}
In summary, the CDFT has been solved in 3D lattice space by implementing the PCG-F method.
It considerably improves the computational efficiency compared to the previous inverse Hamiltonian method. 
Based on this framework, the ground state of $^{110}$Zr has been studied and found to have a tetrahedral shape with $\beta_{20}\approx0.00$, $\beta_{30}\approx0.00$ and $\beta_{32}\neq0.00$.
While it is consistent with the results obtained in the previous MDC-CDFT calculations \cite{Zhao2017PRC}, the present work proves that the tetrahedral ground state of $^{110}$Zr can still exist in a full deformation space but greatly softened.
Specifically, with the inclusion of $\beta_{31}$ and $\beta_{33}$ deformations, the potential energy surface around the tetrahedral minimum becomes much softer in both $\beta_{32}$ and $\beta_{30}$ directions.
The softness of the tetrahedral ground state should be associated with the roughly constant single-particle energy gaps at $N=70$ and $Z=40$.

The softness of the tetrahedral shape due to the $\beta_{31}$ and $\beta_{33}$ deformations might exist in other nuclei as well.
To search for well tetrahedral states in nuclei, the present work has demonstrated the importance of the calculations performed in the full deformation space.
Based on the CDFT in 3D lattice space, works along this line are in progress.

\begin{acknowledgments}
 We thank Y. K. Wang for helpful discussions. This work was partly supported by  the National Natural Science Foundation of China (Grants No. 12070131001, No. 11935003, No. 11975031, and No. 12141501), and the High-performance Computing Platform of Peking University. Z. X. Ren is supported in part by the European Research Council (ERC) under the European Union's Horizon 2020 research and innovation programme (Grant agreement No. 101018170).
\end{acknowledgments}

\end{document}